\documentclass[reprint,superscriptaddress,amssymb,amsmath,aps,showpacs,10pt,floatfix,prl,longbibliography]{revtex4-1}
\def\cm{cm$^{-1}$}
\usepackage{graphicx}%
\usepackage{color}
\usepackage{epstopdf}
\usepackage{amssymb}
\usepackage{amsmath}
\usepackage{amsfonts}
\usepackage{color}
\usepackage[colorlinks,bookmarks=false,citecolor=darkblue,linkcolor=red,urlcolor=blue]{hyperref} 
\definecolor{darkred}{rgb}{0.7,0.0,0.0}

\definecolor{darkblue}{rgb}{0,0.02,0.45}

\definecolor{darkgreen}{rgb}{0.02,0.45,0.0}

\definecolor{violet}{rgb}{0.8,0.2,0.6}

\begin{document}
\title{
Two linear regimes in optical conductivity of a Type-I Weyl semimetal: the case of elemental tellurium\\
}

\author{Diego Rodriguez}
\affiliation{1. Physikalisches Institut, Universit{\"a}t Stuttgart, 70569 Stuttgart, Germany}

\author{Alexander A. Tsirlin}
\email{altsirlin@gmail.com}
\affiliation{Experimental Physics VI, Center for Electronic Correlations and Magnetism, Augsburg University, 86159 Augsburg, Germany}

\author{Tobias Biesner}
\affiliation{1. Physikalisches Institut, Universit{\"a}t Stuttgart, 70569 Stuttgart, Germany}

\author{Teppei Ueno}
\affiliation{Research Institute for Interdisciplinary Science, Okayama University, Okayama, 700-8530, Japan}

\author{Takeshi Takahashi}
\affiliation{Research Institute for Interdisciplinary Science, Okayama University, Okayama, 700-8530, Japan}

\author{Kaya Kobayashi}
\affiliation{Research Institute for Interdisciplinary Science, Okayama University, Okayama, 700-8530, Japan}

\author{Martin Dressel}
\affiliation{1. Physikalisches Institut, Universit{\"a}t Stuttgart, 70569 Stuttgart, Germany}

\author{Ece Uykur}
\email{ece.uykur@pi1.physik.uni-stuttgart.de}
\affiliation{1. Physikalisches Institut, Universit{\"a}t Stuttgart, 70569 Stuttgart, Germany}

\date{\today}

\begin{abstract}

Employing high-pressure infrared spectroscopy we unveil the Weyl semimetal phase of elemental Te and its topological properties. The linear frequency dependence of the optical conductivity provides clear evidence for metallization of trigonal tellurium (Te-I) and the linear band dispersion above 3.0 GPa. This semimetallic Weyl phase can be tuned by increasing pressure further: a kink separates two linear regimes in the optical conductivity (at 3.7 GPa), a signature proposed for Type-II Weyl semimetals with tilted cones; this however reveals a different origin in trigonal tellurium. Our density-functional calculations do not reveal any significant tilting and suggest that Te-I remains in the Type-I Weyl phase, but with two valence bands in the vicinity of the Fermi level. Their interplay gives rise to the peculiar optical conductivity behavior with more than one linear regime. Pressure above 4.3 GPa stabilizes the more complex Te-II and Te-III polymorphs, which are robust metals.

\end{abstract}

\maketitle

Elemental tellurium is known for its unusual properties in the context of efficient thermopower \cite{Lin2016} and possible multiferroic behavior \cite{Furukawa2017}; current research on topological systems brings this material into focus again. Topological insulators, Dirac or Weyl semimetals [(DSMs) or (WSMs)], and three-dimensional topological systems are subject to intense investigations including but not limited to the Te-containing compounds. Especially after the discovery of the three-dimensional topological insulators, which contain heavy group-VI elements (Se and Te), it was proposed that elemental tellurium itself can show intriguing topological properties \cite{Hirayama2015, Agapito2013}; the experimental verification, however, turned out to be challenging.

\begin{figure}
\centering
\includegraphics[width=1\linewidth]{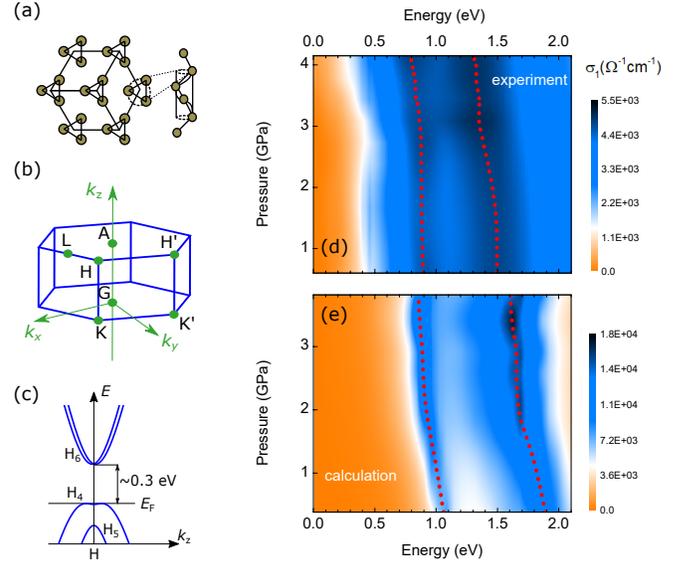}%
\caption{ (a) Crystal structure of trigonal tellurium. (b) Brillouin zone. (c) Band structure around the $H$ point along $k_z$ (d) Pressure-dependent optical conductivity spectra obtained from experiments clearly demonstrate two absorption peaks in the high-energy range. (e) DFT calculations reproduce the positions of the high energy absorption bands and their red shift upon compression. Red dotted lines are the trace of the peak positions showing redshift of the peaks indicating the gradual fill-in of the low energy optical conductivity. }
\label{HE}%
\end{figure}

At ambient conditions, elemental tellurium is a narrow-gap semiconductor. It grows in a trigonal crystal structure [space group P3$_1$21 (right-handed) or P3$_2$21 (left-handed)] with characteristic helical chains containing three atoms in a unit cell [Fig.~\ref{HE}(a)]. These helical chains are arranged in a hexagonal array, resembling graphene, the well-known Dirac semimetal. The intrachain bonds with the nearest-neighbor atoms possess a strong covalent character, while the interchain interactions with the next-nearest neighbor atoms are more of weak van der Waals-type. Based on its space group, trigonal tellurium (Te-I) can be considered as a noncentrosymmetric material, and it is predicted to turn into a topological metal under external pressure and/or uniaxial strain \cite{Hirayama2015, Agapito2013}. The minimum gap at ambient pressure occurs at the $H$ point of the Brillouin zone [Figs.~\ref{HE}(b) and ~\ref{HE}(c)]. Since the $H$ point is not time-reversal invariant, band splitting occurs due to spin-orbit coupling; however, the symmetry of the system requires that two of the bands (conduction bands) are degenerate at the $H$ point forming a Weyl point. Additional Weyl points are expected away from the $H$ point when the system becomes metallic, with valence and conduction bands crossing each other \cite{Hirayama2015, Agapito2013}.

Several computational studies unanimously predict the topological nature of the metallic/ semi\-metallic state in Te-I; the experimental situation, however, is not that simple.
Under external pressure, elemental tellurium goes through a series of structural phase transitions accompanied by a semiconductor-metal transition \cite{Hejny2003, Hejny2004, Marini2012}. It does not become metallic until about 3.5~GPa \cite{Ramasesha1991}; this pressure almost coincides with the structural transformations \cite{Hejny2003, Hejny2004, Marini2012},
which lead to the less studied high-pressure Te-II and Te-III. Since these polymorphs are expected to be metallic, it is plausible that the metallization of tellurium occurs only through the structural transformations. Previous high-pressure optical studies \cite{Yamamoto1995} in fact proposed such a scenario, which is in stark contrast to the predicted pressure-induced topological effects in Te-I. We may note, in passing, that several recent studies explored the effect of Weyl nodes at ambient pressure \cite{Nakayama2017, Zhang2019}, but these putative Weyl nodes lie well below the Fermi level and do not influence the pressure-induced behavior discussed in the following. The aim of the present study is to prove the formation of a metallic state in Te-I and its topologically nontrivial nature that, moreover, appears to be tunable by pressure. To that end, we perform high-pressure infrared spectroscopy on elemental tellurium combined with density-functional (DFT) calculations. Details regarding the experiments and calculations can be found in the Supplemental Material. 

In Fig.~\ref{HE}(d), the room-temperature conductivity $\sigma_1(\omega)$ is plotted in a large frequency range as a function of pressure. The high-energy absorption peaks located at around 0.8 and 1.5~eV at low pressure gradually shift to lower energies; this behavior is accurately reproduced by our DFT calculations [Fig.~\ref{HE}(e)]. At the lowest pressure, a complete gap is observed below ~0.3~eV that gradually fills in upon compression.

\begin{figure}
\centering
\includegraphics[width=1\linewidth]{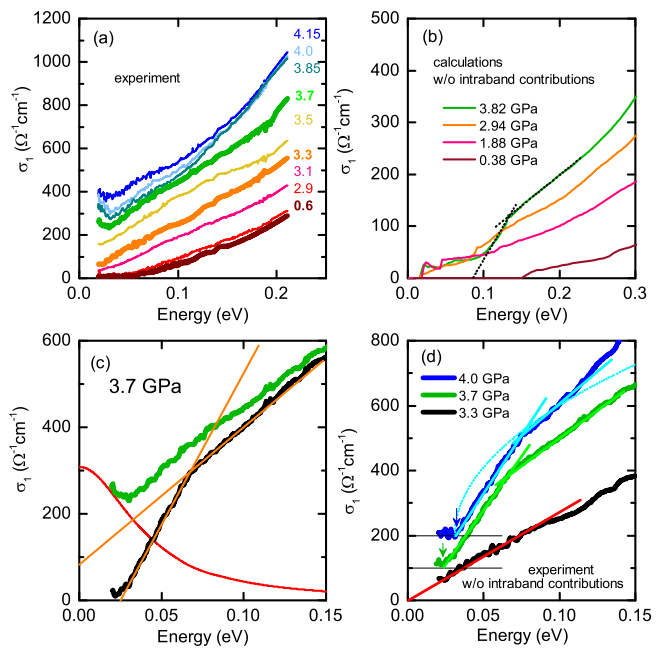}
\caption{ Low-energy optical conductivity in the Te-I phase. (a) Experimental spectra collected at different pressure values as indicated. Thicker lines correspond to the critical pressures, where a change in the profile of the optical conductivity has been observed, ranging from a complete suppression below $\sim$50~meV, to linear-in-energy behavior, and finally two linear regimes separated by a kink. 
(b) Calculated optical conductivity shows a similar evolution upon compression, with the 2.94 GPa curve indicative of the linear-in-energy regime and the 3.82 GPa curve showing two linear regimes as a function of energy. In order to directly compare with the calculations, we subtract the Drude term from the experimental spectra. Panel (c) demonstrates the subtraction procedure on the 3.7 GPa spectrum with (green) and without (black) the intraband contribution (red). (d) Linear-in-frequency optical conductivity at 3.3 GPa evolves to the two linear regimes with increasing pressure. Low-energy optical conductivity does not extrapolate to zero but disappears at a finite energy, also shifts to the higher energies with increasing pressure, indicates that the Weyl points do not lie at the Fermi energy. The spectra are shifted for 3.5 and 4.0~GPa. The dotted blue line is the $\sqrt{\omega}$ fit (expected from parabolic bands) to the 4.0~GPa data for comparison.}
\label{LPLE}%
\end{figure}

The DFT calculations reveal that the band gap at the $H$ point closes when external pressure is applied -- still maintaining the crystal structure of Te-I (see Supplemental Material) -- in full accord with low-energy measurements of the optical conductivity.
In the lower pressure range up to approximately 3~GPa [Fig.~\ref{LPLE}(a)],
the conductivity gradually drops towards low energies and is completely suppressed below 50~meV.
$\sigma_1(\omega)$ assumes a rather linear profile above 3~GPa; the zero-frequency extrapolation strongly indicates that the band gap closes.
When pressure increases further, the low-energy spectral weight rises leading to a finite conductivity in the dc limit around 3.3 to 3.5~GPa. A drastic change in the optical properties becomes obvious above 3.5~GPa. A pronounced Drude-type upturn in $\sigma_1(\omega)$ develops, indicating the significant contribution of the intraband transitions. In addition, at $\sim$3.7~GPa a kinklike structure appears at $\sim$70~meV that persists at high pressures up to 4.3~GPa. This kink is best visible after the subtraction of the intraband contribution [Fig.~\ref{LPLE}(c)], but is quite robust with respect to the subtraction procedure (Supplemental Material)

\begin{figure*}
\centering
\includegraphics[width=1\linewidth]{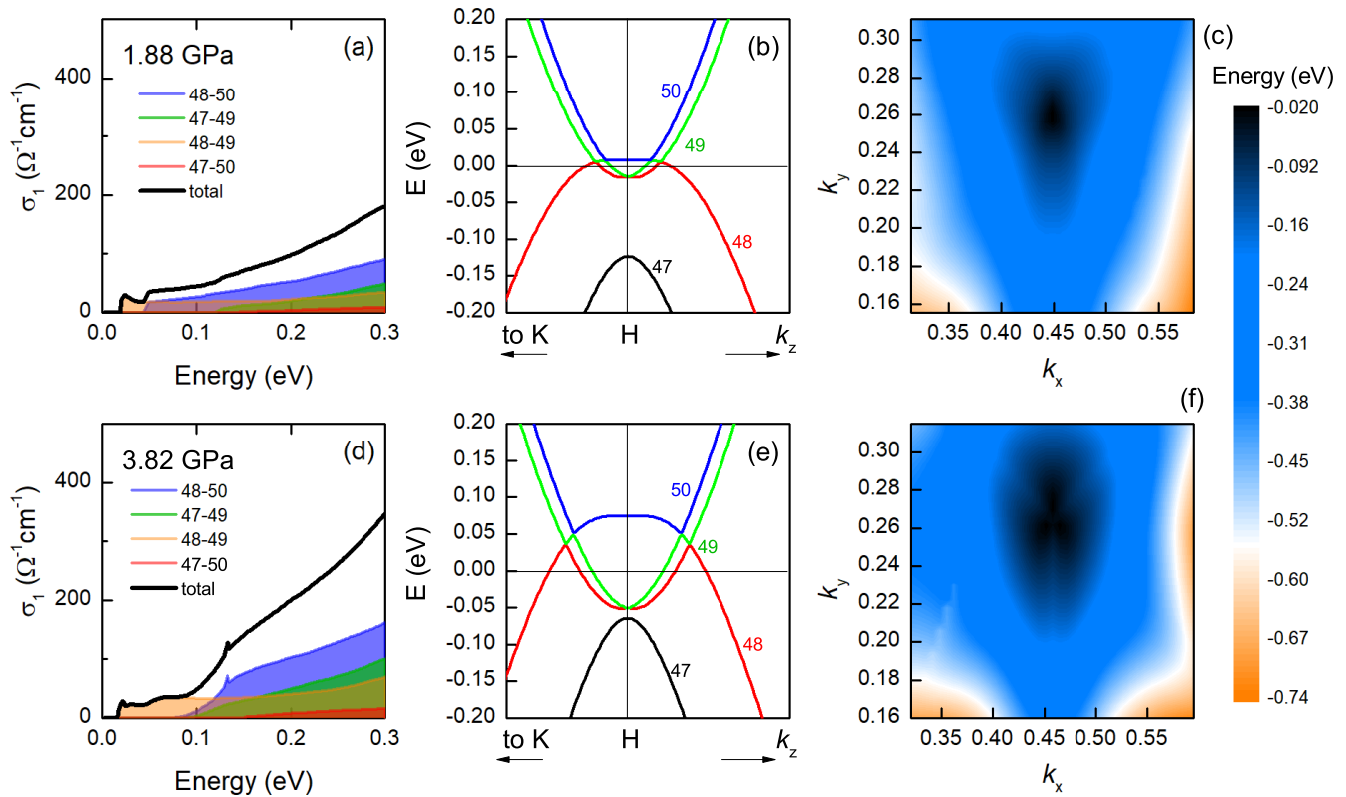}%
\caption{Band-resolved optical conductivity calculations for 1.88 (upper panels) and 3.82~GPa (lower panels). (a and d) calculated optical conductivity and its band-resolved contributions demonstrating the evolution of the kinklike behavior and two linear regimes. This behavior mostly results from the transitions between bands 48 and 50, which also give the dominant contribution to the optical conductivity. (b and e) Band structure around the $H$ point calculated along the $k_z$ direction. Optical conductivities are calculated by taking into account the corresponding transitions between the shown bands. (c and f) 2D profile of the band 48 along the $k_x$ and $k_z$ directions demonstrating the band flattening around the $H$ point upon increasing pressure from 1.88~GPa to 3.82~GPa. $k_x$ and $k_y$ are given in the Cartesian frame in units of the reciprocal lattice parameter $a^*=2\pi/(a\sin\gamma)$ with $\gamma=120^{\circ}$. Other bands around the $H$ point show a minor change only, see the Supplemental Material. }
\label{calc}%
\end{figure*}

\begin{figure}
\centering
\includegraphics[width=1\linewidth]{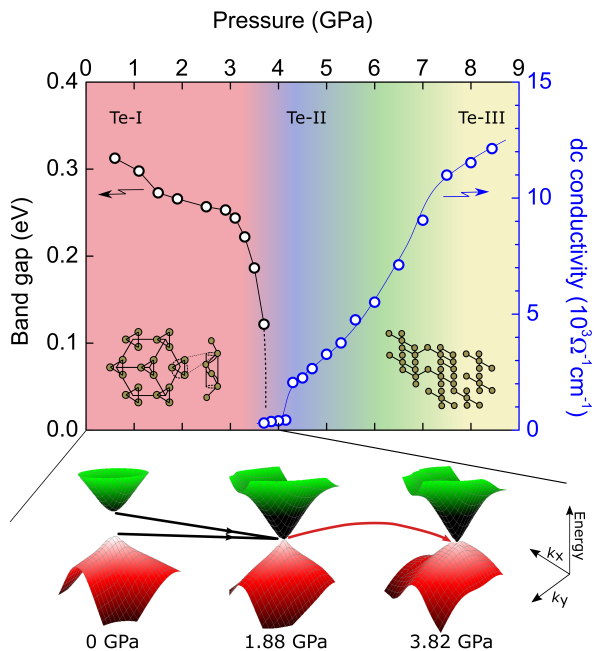}%
\caption{The high pressure phase diagram of elemental tellurium based on the infrared studies and corroborated by high-pressure x-ray diffraction studies \cite{Hejny2003, Hejny2004}.  Lower panels show the upper valence band and the lower conduction band of Te-I at different pressures at the edge of the Brillouin zone ($k_z=0.5$) around the $H$ point. The evolution of these bands in the Te-I phase is demonstrated with the arrows: While the gap closes gradually and band touching creates linear band dispersion at 1.88~GPa (black arrows), the valence band is strongly modified with further increasing pressure and flattens developing three local maxima (red arrow).  }
\label{PD}%
\end{figure}

Already from these observations we conclude that the metallization of Te occurs independently from and prior to the structural phase transition. In fact, diffraction data first see the high-pressure Te-II polymorph only above 3.8~GPa \cite{Hejny2003, Hejny2004}, which is at least 0.3~GPa above the pressure where $\sigma_1(\omega)$ extrapolates to a nonzero dc value, while accuracy of our pressure values is 0.15~GPa (Supplemental Material). These findings are also in line with the rather low dc conductivity observed in our data up to 4.3~GPa [see Fig.~\ref{PD}] indicating the dominant contribution of the small Fermi surface of the Te-I phase that persists up to 4.3~GPa.

The linear-in-energy behavior of $\sigma_1(\omega)$  observed above 3.1~GPa is very peculiar, especially in the context of the topologically nontrivial states. The linear band dispersion is the hallmark of the Dirac and Weyl semimetals and gives rise to a linear optical conductivity that is observed in various 3D systems \cite{Hosur2012, Tabert2016}. More generally, the power-law behavior of the band dispersion, $E(k) \propto k^z$, is reflected in the real part of the optical conductivity via $\sigma_1(\omega)\propto\omega^{(d-2)/z}$. Here, $d$ is the dimension of the system; i.e.\ for a linear band dispersion in $d=3$ one finds $\sigma_1(\omega)\propto\omega$, the celebrated linear regime of $\sigma_1(\omega)$ for the three-dimensional systems with linear band dispersion \cite{Xu2016, Kimura2017}.  When a gap opening occurs at the Dirac point and/or the shift of the chemical potential with respect to the Dirac nodes needs to be considered, modifications to this most general case are necessary \cite{Ashby2014, Tabert2016, Neubauer2016, Neubauer2018, Huett2018}.

Pressure evolution of the low-frequency conductivity is mirrored by our DFT calculations [Fig.~\ref{LPLE}(b)]. Gradually, the low-energy spectral weight decreases until it is completely suppressed and the linear-in-energy behavior becomes obvious; two linear regimes separated by a kink at 3.82~GPa can be identified. This strongly suggests that the linear optical conductivity appears when the gap closes at the $H$ point and that the ensuing metallic state evolves upon further compression. Combining these calculations with our experiments, we clearly conclude the existence of a metallic Te-I phase with linear band dispersion; hence this electronic transition is topological in nature.

Let us now turn to the kinklike optical conductivity present above 3.7~GPa. A closer look at the low-frequency region [Fig.~\ref{LPLE}(d)] reveals that $\sigma_1(\omega)$ does not extrapolate to zero but disappears around 25~meV at 3.7 GPa and blueshifts with increasing pressure [arrows in Fig.~\ref{LPLE}(d) demonstrate the shift] indicating that the Weyl points do not lie at the Fermi level, but are shifted in energy. This is in stark contrast to the behavior observed at 3.3~GPa, when the bands just touch each other; in that case the linear regime extrapolates to zero frequency indicating that the Weyl points actually fall right onto the Fermi level. In other words, the results from our optical studies evidence that the Weyl points move away from the Fermi energy with increasing pressure in full accord with our DFT calculations [see Figs.~\ref{calc}(b) and ~\ref{calc}(e)].

The kinklike behavior above 3.7~GPa is peculiar because a recent theoretical study predicts the two linear regimes with a kink in the optical conductivity arising from the tilt of the Weyl cones \cite{Carbotte2016}, e.g.\ in type-II WSMs, such as MoTe$_2$ \cite{Jiang2017, santos2019, Kimura2019}, WTe$_2$ \cite{Homes2015, Wang2016, Frenzel2017, Kimura2019}. Hence, it is very tempting to interpret our optical conductivity within the same framework, especially in light of Te being present in the aforementioned compounds. However, the prediction of the two linear regimes due to the tilted Weyl cones crucially relies on the presence of only two linear bands around the Fermi level, which is not the case for Te-I [Figs.~\ref{calc}(b) and ~\ref{calc}(e)].

To elucidate the origin of the two linear regimes in the optical conductivity of Te-I, we performed band-resolved calculations. The results for two particular situations are displayed in Fig.~\ref{calc}. The upper panels are taken at 1.88~GPa when the bands just touch each other, while the lower panels are at 3.82 GPa, deep in the metallic phase; bands are numbered from bottom to top. The largest contribution to the low-energy conductivity is given by the transition between bands 48 and 50.
As pressure increases further, the most obvious change occurs in the band 48 along all momentum directions [Figs.~\ref{calc}(c) and ~\ref{calc}(f)]: the band strongly flattens around its maximum.
A closer look unveils that at $p=3.82$~GPa the band maximum actually consists of three band maxima.
Considering that the kinklike feature is prominently seen in the transitions between bands 48 and 50, we attribute it to the change in the structure of band 48. A detailed comparison for the pressure evolution of other bands is given in Supplemental Material. Note, there is no significant tilt around the band crossing points; minor tilts, however, are not sufficient to account for the two linear regimes of the experimental data.
This implies that the two linear regimes in $\sigma_1(\omega)$ have a very different origin: they arise from changes in band 48 that, in turn, is affected by band 47, which rises in energy as pressure increases [Fig.~\ref{calc}(e)].
This scenario is essentially dissimilar to the two linear bands considered in model studies  \cite{Carbotte2016}.
The situation is also very distinct from the one observed in Te-based topological materials such as BiTeI \cite{Xi2013, Tran2014}.

Our optical results clearly demonstrate the topologically nontrivial metallic state in Te-I,
which is characterized by the low carrier density and small dc conductivity. 
This changes drastically above 4.3~GPa, where the system enters Te-II and Te-III phases. Concordantly the optical spectra are significantly modified. The Te-I to Te-II transition manifests itself in a sudden jump of the dc conductivity [Fig. ~\ref{PD}] at 4.3~GPa, indicating a drastic change of the electronic structure, while the changes through the Te-III phase are more subtle owing to the fact that Te-II and Te-III phases are structurally related (pressure-dependent reflectivity of the Te-II and Te-III are given in the Supplemental Material). The signatures of the topologically nontrivial phase do not survive upon the structural transitions, therefore the high pressure polymorphs are beyond the scope of our present work.

In summary, by applying broadband infrared measurements on elemental tellurium under external pressure, we resolved the evolution of its electronic structure within the trigonal Te-I phase and further across the pressure-induced structural phase transitions.
The formation of metallic Te-I above 3.1~GPa and the topological nature of this metallic phase are confirmed. Moreover, tunability of the topological phase within the 3.1 - 4.3~GPa pressure range is demonstrated. Our work not only sheds light onto the topological properties of a simple elemental solid, but also raises an important caveat for the optical signatures of putative Type-II Weyl states in a multiband systems.

\begin{acknowledgments}
We acknowledge fruitful discussions with Artem V. Pronin and Sascha Polatkan; technical support from Gabriele Untereiner. We are grateful to Malcolm McMahon for sharing his x-ray data for our calculations. Work in Okayama is supported by the Grant-in-Aid for Scientific Research (Grant Numbers: 18K03540, 19H01852). E.U. acknowledges the European Social Fund and the Baden-Württemberg Stiftung for the financial support of this research project by the Eliteprogramme.
\end{acknowledgments}

\bibliography{MainTe}

\begin{thebibliography}{28}%
\makeatletter
\providecommand \@ifxundefined [1]{%
 \@ifx{#1\undefined}
}%
\providecommand \@ifnum [1]{%
 \ifnum #1\expandafter \@firstoftwo
 \else \expandafter \@secondoftwo
 \fi
}%
\providecommand \@ifx [1]{%
 \ifx #1\expandafter \@firstoftwo
 \else \expandafter \@secondoftwo
 \fi
}%
\providecommand \natexlab [1]{#1}%
\providecommand \enquote  [1]{``#1''}%
\providecommand \bibnamefont  [1]{#1}%
\providecommand \bibfnamefont [1]{#1}%
\providecommand \citenamefont [1]{#1}%
\providecommand \href@noop [0]{\@secondoftwo}%
\providecommand \href [0]{\begingroup \@sanitize@url \@href}%
\providecommand \@href[1]{\@@startlink{#1}\@@href}%
\providecommand \@@href[1]{\endgroup#1\@@endlink}%
\providecommand \@sanitize@url [0]{\catcode `\\12\catcode `\$12\catcode
  `\&12\catcode `\#12\catcode `\^12\catcode `\_12\catcode `\%12\relax}%
\providecommand \@@startlink[1]{}%
\providecommand \@@endlink[0]{}%
\providecommand \url  [0]{\begingroup\@sanitize@url \@url }%
\providecommand \@url [1]{\endgroup\@href {#1}{\urlprefix }}%
\providecommand \urlprefix  [0]{URL }%
\providecommand \Eprint [0]{\href }%
\providecommand \doibase [0]{https://doi.org/}%
\providecommand \selectlanguage [0]{\@gobble}%
\providecommand \bibinfo  [0]{\@secondoftwo}%
\providecommand \bibfield  [0]{\@secondoftwo}%
\providecommand \translation [1]{[#1]}%
\providecommand \BibitemOpen [0]{}%
\providecommand \bibitemStop [0]{}%
\providecommand \bibitemNoStop [0]{.\EOS\space}%
\providecommand \EOS [0]{\spacefactor3000\relax}%
\providecommand \BibitemShut  [1]{\csname bibitem#1\endcsname}%
\let\auto@bib@innerbib\@empty
%</preamble>
\bibitem [{\citenamefont {Lin}\ \emph {et~al.}(2016)\citenamefont {Lin},
  \citenamefont {Li}, \citenamefont {Chen}, \citenamefont {Shen}, \citenamefont
  {Ge},\ and\ \citenamefont {Pei}}]{Lin2016}%
  \BibitemOpen
  \bibfield  {author} {\bibinfo {author} {\bibfnamefont {S.}~\bibnamefont
  {Lin}}, \bibinfo {author} {\bibfnamefont {W.}~\bibnamefont {Li}}, \bibinfo
  {author} {\bibfnamefont {Z.}~\bibnamefont {Chen}}, \bibinfo {author}
  {\bibfnamefont {J.}~\bibnamefont {Shen}}, \bibinfo {author} {\bibfnamefont
  {B.}~\bibnamefont {Ge}},\ and\ \bibinfo {author} {\bibfnamefont
  {Y.}~\bibnamefont {Pei}},\ }\bibfield  {title} {\bibinfo {title} {{Tellurium
  as a high-performance elemental thermoelectric}},\ }\href
  {https://doi.org/https://doi.org/10.1038/ncomms10287} {\bibfield  {journal}
  {\bibinfo  {journal} {Nature Communications}\ }\textbf {\bibinfo {volume}
  {7}},\ \bibinfo {pages} {10287} (\bibinfo {year} {2016})}\BibitemShut
  {NoStop}%
\bibitem [{\citenamefont {Furukawa}\ \emph {et~al.}(2017)\citenamefont
  {Furukawa}, \citenamefont {Shimokawa}, \citenamefont {Kobayashi},\ and\
  \citenamefont {Itou}}]{Furukawa2017}%
  \BibitemOpen
  \bibfield  {author} {\bibinfo {author} {\bibfnamefont {T.}~\bibnamefont
  {Furukawa}}, \bibinfo {author} {\bibfnamefont {Y.}~\bibnamefont {Shimokawa}},
  \bibinfo {author} {\bibfnamefont {K.}~\bibnamefont {Kobayashi}},\ and\
  \bibinfo {author} {\bibfnamefont {T.}~\bibnamefont {Itou}},\ }\bibfield
  {title} {\bibinfo {title} {{Observation of current-induced bulk magnetization
  in elemental tellurium}},\ }\href
  {https://doi.org/https://doi.org/10.1038/s41467-017-01093-3} {\bibfield
  {journal} {\bibinfo  {journal} {Nature Communications}\ }\textbf {\bibinfo
  {volume} {8}},\ \bibinfo {pages} {954} (\bibinfo {year} {2017})}\BibitemShut
  {NoStop}%
\bibitem [{\citenamefont {Hirayama}\ \emph {et~al.}(2015)\citenamefont
  {Hirayama}, \citenamefont {Okugawa}, \citenamefont {Ishibashi}, \citenamefont
  {Murakami},\ and\ \citenamefont {Miyake}}]{Hirayama2015}%
  \BibitemOpen
  \bibfield  {author} {\bibinfo {author} {\bibfnamefont {M.}~\bibnamefont
  {Hirayama}}, \bibinfo {author} {\bibfnamefont {R.}~\bibnamefont {Okugawa}},
  \bibinfo {author} {\bibfnamefont {S.}~\bibnamefont {Ishibashi}}, \bibinfo
  {author} {\bibfnamefont {S.}~\bibnamefont {Murakami}},\ and\ \bibinfo
  {author} {\bibfnamefont {T.}~\bibnamefont {Miyake}},\ }\bibfield  {title}
  {\bibinfo {title} {{Weyl Node and Spin Texture in Trigonal Tellurium and
  Selenium}},\ }\href {https://doi.org/10.1103/PhysRevLett.114.206401}
  {\bibfield  {journal} {\bibinfo  {journal} {Phys. Rev. Lett.}\ }\textbf
  {\bibinfo {volume} {114}},\ \bibinfo {pages} {206401} (\bibinfo {year}
  {2015})}\BibitemShut {NoStop}%
\bibitem [{\citenamefont {Agapito}\ \emph {et~al.}(2013)\citenamefont
  {Agapito}, \citenamefont {Kioussis}, \citenamefont {Goddard},\ and\
  \citenamefont {Ong}}]{Agapito2013}%
  \BibitemOpen
  \bibfield  {author} {\bibinfo {author} {\bibfnamefont {L.~A.}\ \bibnamefont
  {Agapito}}, \bibinfo {author} {\bibfnamefont {N.}~\bibnamefont {Kioussis}},
  \bibinfo {author} {\bibfnamefont {W.~A.}\ \bibnamefont {Goddard}},\ and\
  \bibinfo {author} {\bibfnamefont {N.~P.}\ \bibnamefont {Ong}},\ }\bibfield
  {title} {\bibinfo {title} {{Novel Family of Chiral-Based Topological
  Insulators: Elemental Tellurium under Strain}},\ }\href
  {https://doi.org/10.1103/PhysRevLett.110.176401} {\bibfield  {journal}
  {\bibinfo  {journal} {Phys. Rev. Lett.}\ }\textbf {\bibinfo {volume} {110}},\
  \bibinfo {pages} {176401} (\bibinfo {year} {2013})}\BibitemShut {NoStop}%
\bibitem [{\citenamefont {Hejny}\ and\ \citenamefont
  {McMahon}(2003)}]{Hejny2003}%
  \BibitemOpen
  \bibfield  {author} {\bibinfo {author} {\bibfnamefont {C.}~\bibnamefont
  {Hejny}}\ and\ \bibinfo {author} {\bibfnamefont {M.}~\bibnamefont
  {McMahon}},\ }\bibfield  {title} {\bibinfo {title} {{Large Structural
  Modulations in Incommensurate {Te-III} and {Se-IV}}},\ }\href
  {https://doi.org/10.1103/PhysRevLett.91.215502} {\bibfield  {journal}
  {\bibinfo  {journal} {Phys. Rev. Lett.}\ }\textbf {\bibinfo {volume} {91}},\
  \bibinfo {pages} {215502} (\bibinfo {year} {2003})}\BibitemShut {NoStop}%
\bibitem [{\citenamefont {Hejny}\ and\ \citenamefont
  {McMahon}(2004)}]{Hejny2004}%
  \BibitemOpen
  \bibfield  {author} {\bibinfo {author} {\bibfnamefont {C.}~\bibnamefont
  {Hejny}}\ and\ \bibinfo {author} {\bibfnamefont {M.}~\bibnamefont
  {McMahon}},\ }\bibfield  {title} {\bibinfo {title} {{Complex crystal
  structures of {Te-II} and {Se-III} at high pressure}},\ }\href
  {https://doi.org/10.1103/PhysRevB.70.184109} {\bibfield  {journal} {\bibinfo
  {journal} {Phys. Rev. B}\ }\textbf {\bibinfo {volume} {70}},\ \bibinfo
  {pages} {184109} (\bibinfo {year} {2004})}\BibitemShut {NoStop}%
\bibitem [{\citenamefont {Marini}\ \emph {et~al.}(2012)\citenamefont {Marini},
  \citenamefont {Chermisi}, \citenamefont {Lavagnini}, \citenamefont
  {Di~Castro}, \citenamefont {Petrillo}, \citenamefont {Degiorgi},
  \citenamefont {Scandolo},\ and\ \citenamefont {Postorino}}]{Marini2012}%
  \BibitemOpen
  \bibfield  {author} {\bibinfo {author} {\bibfnamefont {C.}~\bibnamefont
  {Marini}}, \bibinfo {author} {\bibfnamefont {D.}~\bibnamefont {Chermisi}},
  \bibinfo {author} {\bibfnamefont {M.}~\bibnamefont {Lavagnini}}, \bibinfo
  {author} {\bibfnamefont {D.}~\bibnamefont {Di~Castro}}, \bibinfo {author}
  {\bibfnamefont {C.}~\bibnamefont {Petrillo}}, \bibinfo {author}
  {\bibfnamefont {L.}~\bibnamefont {Degiorgi}}, \bibinfo {author}
  {\bibfnamefont {S.}~\bibnamefont {Scandolo}},\ and\ \bibinfo {author}
  {\bibfnamefont {P.}~\bibnamefont {Postorino}},\ }\bibfield  {title} {\bibinfo
  {title} {High-pressure phases of crystalline tellurium: A combined raman and
  ab initio study},\ }\href
  {https://doi.org/https://doi.org/10.1103/PhysRevB.86.064103} {\bibfield
  {journal} {\bibinfo  {journal} {Phys. Rev. B}\ }\textbf {\bibinfo {volume}
  {86}},\ \bibinfo {pages} {064103} (\bibinfo {year} {2012})}\BibitemShut
  {NoStop}%
\bibitem [{\citenamefont {Ramasesha}\ and\ \citenamefont
  {Singh}(1991)}]{Ramasesha1991}%
  \BibitemOpen
  \bibfield  {author} {\bibinfo {author} {\bibfnamefont {S.~K.}\ \bibnamefont
  {Ramasesha}}\ and\ \bibinfo {author} {\bibfnamefont {A.~K.}\ \bibnamefont
  {Singh}},\ }\bibfield  {title} {\bibinfo {title} {{Thermoelectric power of
  tellurium under pressure up to 8 GPa}},\ }\href
  {https://doi.org/https://doi.org/10.1080/13642819108217880} {\bibfield
  {journal} {\bibinfo  {journal} {Philosophical Magazine B}\ }\textbf {\bibinfo
  {volume} {64}},\ \bibinfo {pages} {559} (\bibinfo {year} {1991})}\BibitemShut
  {NoStop}%
\bibitem [{\citenamefont {Yamamoto}\ \emph {et~al.}(1995)\citenamefont
  {Yamamoto}, \citenamefont {Ohmasa}, \citenamefont {Ikeda},\ and\
  \citenamefont {Endo}}]{Yamamoto1995}%
  \BibitemOpen
  \bibfield  {author} {\bibinfo {author} {\bibfnamefont {I.}~\bibnamefont
  {Yamamoto}}, \bibinfo {author} {\bibfnamefont {Y.}~\bibnamefont {Ohmasa}},
  \bibinfo {author} {\bibfnamefont {H.}~\bibnamefont {Ikeda}},\ and\ \bibinfo
  {author} {\bibfnamefont {H.}~\bibnamefont {Endo}},\ }\bibfield  {title}
  {\bibinfo {title} {{The optical properties of tellurium under high
  pressure}},\ }\href {https://doi.org/10.1088/0953-8984/7/22/012} {\bibfield
  {journal} {\bibinfo  {journal} {Journal of Physics: Condensed Matter}\
  }\textbf {\bibinfo {volume} {7}},\ \bibinfo {pages} {4299} (\bibinfo {year}
  {1995})}\BibitemShut {NoStop}%
\bibitem [{\citenamefont {Nakayama}\ \emph {et~al.}(2017)\citenamefont
  {Nakayama}, \citenamefont {Kuno}, \citenamefont {Yamauchi}, \citenamefont
  {Souma}, \citenamefont {Sugawara}, \citenamefont {Oguchi}, \citenamefont
  {Sato},\ and\ \citenamefont {Takahashi}}]{Nakayama2017}%
  \BibitemOpen
  \bibfield  {author} {\bibinfo {author} {\bibfnamefont {K.}~\bibnamefont
  {Nakayama}}, \bibinfo {author} {\bibfnamefont {M.}~\bibnamefont {Kuno}},
  \bibinfo {author} {\bibfnamefont {K.}~\bibnamefont {Yamauchi}}, \bibinfo
  {author} {\bibfnamefont {S.}~\bibnamefont {Souma}}, \bibinfo {author}
  {\bibfnamefont {K.}~\bibnamefont {Sugawara}}, \bibinfo {author}
  {\bibfnamefont {T.}~\bibnamefont {Oguchi}}, \bibinfo {author} {\bibfnamefont
  {T.}~\bibnamefont {Sato}},\ and\ \bibinfo {author} {\bibfnamefont
  {T.}~\bibnamefont {Takahashi}},\ }\bibfield  {title} {\bibinfo {title} {{Band
  splitting and Weyl nodes in trigonal tellurium studied by angle-resolved
  photoemission spectroscopy and density functional theory}},\ }\href
  {https://doi.org/https://doi.org/10.1103/PhysRevB.95.125204} {\bibfield
  {journal} {\bibinfo  {journal} {Phys. Rev. B}\ }\textbf {\bibinfo {volume}
  {95}},\ \bibinfo {pages} {125204} (\bibinfo {year} {2017})}\BibitemShut
  {NoStop}%
\bibitem [{\citenamefont {Zhang}\ \emph {et~al.}(2019)\citenamefont {Zhang},
  \citenamefont {Zhao}, \citenamefont {Li}, \citenamefont {Wang}, \citenamefont
  {Xie}, \citenamefont {Li}, \citenamefont {Lin}, \citenamefont {He},
  \citenamefont {Sun}, \citenamefont {Wang}, \citenamefont {Zhang},\ and\
  \citenamefont {Zeng}}]{Zhang2019}%
  \BibitemOpen
  \bibfield  {author} {\bibinfo {author} {\bibfnamefont {N.}~\bibnamefont
  {Zhang}}, \bibinfo {author} {\bibfnamefont {G.}~\bibnamefont {Zhao}},
  \bibinfo {author} {\bibfnamefont {L.}~\bibnamefont {Li}}, \bibinfo {author}
  {\bibfnamefont {P.}~\bibnamefont {Wang}}, \bibinfo {author} {\bibfnamefont
  {L.}~\bibnamefont {Xie}}, \bibinfo {author} {\bibfnamefont {H.}~\bibnamefont
  {Li}}, \bibinfo {author} {\bibfnamefont {Z.}~\bibnamefont {Lin}}, \bibinfo
  {author} {\bibfnamefont {J.}~\bibnamefont {He}}, \bibinfo {author}
  {\bibfnamefont {Z.}~\bibnamefont {Sun}}, \bibinfo {author} {\bibfnamefont
  {Z.}~\bibnamefont {Wang}}, \bibinfo {author} {\bibfnamefont {Z.}~\bibnamefont
  {Zhang}},\ and\ \bibinfo {author} {\bibfnamefont {C.}~\bibnamefont {Zeng}},\
  }\bibfield  {title} {\bibinfo {title} {{Evidence for Weyl fermions in the
  elemental semiconductor tellurium}},\ }\href@noop {} {\bibfield  {journal}
  {\bibinfo  {journal} {ARXIV: 1906.06071}\ } (\bibinfo {year}
  {2019})}\BibitemShut {NoStop}%
\bibitem [{\citenamefont {Hosur}\ \emph {et~al.}(2012)\citenamefont {Hosur},
  \citenamefont {Parameswaran},\ and\ \citenamefont {Vishwanath}}]{Hosur2012}%
  \BibitemOpen
  \bibfield  {author} {\bibinfo {author} {\bibfnamefont {P.}~\bibnamefont
  {Hosur}}, \bibinfo {author} {\bibfnamefont {S.~A.}\ \bibnamefont
  {Parameswaran}},\ and\ \bibinfo {author} {\bibfnamefont {A.}~\bibnamefont
  {Vishwanath}},\ }\bibfield  {title} {\bibinfo {title} {{Charge Transport in
  Weyl Semimetals}},\ }\href
  {https://doi.org/https://doi.org/10.1103/PhysRevLett.108.046602} {\bibfield
  {journal} {\bibinfo  {journal} {Phys. Rev. Lett.}\ }\textbf {\bibinfo
  {volume} {108}},\ \bibinfo {pages} {046602} (\bibinfo {year}
  {2012})}\BibitemShut {NoStop}%
\bibitem [{\citenamefont {Tabert}\ \emph {et~al.}(2016)\citenamefont {Tabert},
  \citenamefont {Carbotte},\ and\ \citenamefont {Nicol}}]{Tabert2016}%
  \BibitemOpen
  \bibfield  {author} {\bibinfo {author} {\bibfnamefont {C.~J.}\ \bibnamefont
  {Tabert}}, \bibinfo {author} {\bibfnamefont {J.~P.}\ \bibnamefont
  {Carbotte}},\ and\ \bibinfo {author} {\bibfnamefont {E.~J.}\ \bibnamefont
  {Nicol}},\ }\bibfield  {title} {\bibinfo {title} {{Optical and transport
  properties in three-dimensional Dirac and Weyl semimetals}},\ }\href
  {https://doi.org/10.1103/PhysRevB.93.085426} {\bibfield  {journal} {\bibinfo
  {journal} {Phys. Rev. B}\ }\textbf {\bibinfo {volume} {93}},\ \bibinfo
  {pages} {085426} (\bibinfo {year} {2016})}\BibitemShut {NoStop}%
\bibitem [{\citenamefont {Xu}\ \emph {et~al.}(2016)\citenamefont {Xu},
  \citenamefont {Dai}, \citenamefont {Zhao}, \citenamefont {Wang},
  \citenamefont {Yang}, \citenamefont {Zhang}, \citenamefont {Liu},
  \citenamefont {Xiao}, \citenamefont {Chen}, \citenamefont {Taylor},
  \citenamefont {Yarotski}, \citenamefont {Prasankumar},\ and\ \citenamefont
  {Qiu}}]{Xu2016}%
  \BibitemOpen
  \bibfield  {author} {\bibinfo {author} {\bibfnamefont {B.}~\bibnamefont
  {Xu}}, \bibinfo {author} {\bibfnamefont {Y.~M.}\ \bibnamefont {Dai}},
  \bibinfo {author} {\bibfnamefont {L.~X.}\ \bibnamefont {Zhao}}, \bibinfo
  {author} {\bibfnamefont {K.}~\bibnamefont {Wang}}, \bibinfo {author}
  {\bibfnamefont {R.}~\bibnamefont {Yang}}, \bibinfo {author} {\bibfnamefont
  {W.}~\bibnamefont {Zhang}}, \bibinfo {author} {\bibfnamefont {J.~Y.}\
  \bibnamefont {Liu}}, \bibinfo {author} {\bibfnamefont {H.}~\bibnamefont
  {Xiao}}, \bibinfo {author} {\bibfnamefont {G.~F.}\ \bibnamefont {Chen}},
  \bibinfo {author} {\bibfnamefont {A.~J.}\ \bibnamefont {Taylor}}, \bibinfo
  {author} {\bibfnamefont {D.~A.}\ \bibnamefont {Yarotski}}, \bibinfo {author}
  {\bibfnamefont {R.~P.}\ \bibnamefont {Prasankumar}},\ and\ \bibinfo {author}
  {\bibfnamefont {X.~G.}\ \bibnamefont {Qiu}},\ }\bibfield  {title} {\bibinfo
  {title} {{Optical spectroscopy of the Weyl semimetal TaAs}},\ }\href
  {https://doi.org/10.1103/PhysRevB.93.121110} {\bibfield  {journal} {\bibinfo
  {journal} {Phys. Rev. B}\ }\textbf {\bibinfo {volume} {93}},\ \bibinfo
  {pages} {121110} (\bibinfo {year} {2016})}\BibitemShut {NoStop}%
\bibitem [{\citenamefont {Kimura}\ \emph {et~al.}(2017)\citenamefont {Kimura},
  \citenamefont {Yokoyama}, \citenamefont {Watanabe}, \citenamefont
  {Sichelschmidt}, \citenamefont {S\"u\ss{}}, \citenamefont {Schmidt},\ and\
  \citenamefont {Felser}}]{Kimura2017}%
  \BibitemOpen
  \bibfield  {author} {\bibinfo {author} {\bibfnamefont {S.-i.}\ \bibnamefont
  {Kimura}}, \bibinfo {author} {\bibfnamefont {H.}~\bibnamefont {Yokoyama}},
  \bibinfo {author} {\bibfnamefont {H.}~\bibnamefont {Watanabe}}, \bibinfo
  {author} {\bibfnamefont {J.}~\bibnamefont {Sichelschmidt}}, \bibinfo {author}
  {\bibfnamefont {V.}~\bibnamefont {S\"u\ss{}}}, \bibinfo {author}
  {\bibfnamefont {M.}~\bibnamefont {Schmidt}},\ and\ \bibinfo {author}
  {\bibfnamefont {C.}~\bibnamefont {Felser}},\ }\bibfield  {title} {\bibinfo
  {title} {{Optical signature of Weyl electronic structures in tantalum
  pnictides $\mathrm{Ta}Pn$ ($Pn=$ P, As)}},\ }\href
  {https://doi.org/10.1103/PhysRevB.96.075119} {\bibfield  {journal} {\bibinfo
  {journal} {Phys. Rev. B}\ }\textbf {\bibinfo {volume} {96}},\ \bibinfo
  {pages} {075119} (\bibinfo {year} {2017})}\BibitemShut {NoStop}%
\bibitem [{\citenamefont {Ashby}\ and\ \citenamefont
  {Carbotte}(2014)}]{Ashby2014}%
  \BibitemOpen
  \bibfield  {author} {\bibinfo {author} {\bibfnamefont {P.}~\bibnamefont
  {Ashby}}\ and\ \bibinfo {author} {\bibfnamefont {J.}~\bibnamefont
  {Carbotte}},\ }\bibfield  {title} {\bibinfo {title} {{Chiral anomaly and
  optical absorption in Weyl semimetals}},\ }\href
  {https://doi.org/https://doi.org/10.1103/PhysRevB.89.245121} {\bibfield
  {journal} {\bibinfo  {journal} {Phys. Rev. B}\ }\textbf {\bibinfo {volume}
  {89}},\ \bibinfo {pages} {245121} (\bibinfo {year} {2014})}\BibitemShut
  {NoStop}%
\bibitem [{\citenamefont {Neubauer}\ \emph {et~al.}(2016)\citenamefont
  {Neubauer}, \citenamefont {Carbotte}, \citenamefont {Nateprov}, \citenamefont
  {L\"ohle}, \citenamefont {Dressel},\ and\ \citenamefont
  {Pronin}}]{Neubauer2016}%
  \BibitemOpen
  \bibfield  {author} {\bibinfo {author} {\bibfnamefont {D.}~\bibnamefont
  {Neubauer}}, \bibinfo {author} {\bibfnamefont {J.~P.}\ \bibnamefont
  {Carbotte}}, \bibinfo {author} {\bibfnamefont {A.~A.}\ \bibnamefont
  {Nateprov}}, \bibinfo {author} {\bibfnamefont {A.}~\bibnamefont {L\"ohle}},
  \bibinfo {author} {\bibfnamefont {M.}~\bibnamefont {Dressel}},\ and\ \bibinfo
  {author} {\bibfnamefont {A.~V.}\ \bibnamefont {Pronin}},\ }\bibfield  {title}
  {\bibinfo {title} {{Interband optical conductivity of the [001]-oriented
  Dirac semimetal ${\mathrm{Cd}}_{3}{\mathrm{As}}_{2}$}},\ }\href
  {https://doi.org/10.1103/PhysRevB.93.121202} {\bibfield  {journal} {\bibinfo
  {journal} {Phys. Rev. B}\ }\textbf {\bibinfo {volume} {93}},\ \bibinfo
  {pages} {121202} (\bibinfo {year} {2016})}\BibitemShut {NoStop}%
\bibitem [{\citenamefont {Neubauer}\ \emph {et~al.}(2018)\citenamefont
  {Neubauer}, \citenamefont {Yaresko}, \citenamefont {Li}, \citenamefont
  {L\"ohle}, \citenamefont {H\"ubner}, \citenamefont {Schilling}, \citenamefont
  {Shekhar}, \citenamefont {Felser}, \citenamefont {Dressel},\ and\
  \citenamefont {Pronin}}]{Neubauer2018}%
  \BibitemOpen
  \bibfield  {author} {\bibinfo {author} {\bibfnamefont {D.}~\bibnamefont
  {Neubauer}}, \bibinfo {author} {\bibfnamefont {A.}~\bibnamefont {Yaresko}},
  \bibinfo {author} {\bibfnamefont {W.}~\bibnamefont {Li}}, \bibinfo {author}
  {\bibfnamefont {A.}~\bibnamefont {L\"ohle}}, \bibinfo {author} {\bibfnamefont
  {R.}~\bibnamefont {H\"ubner}}, \bibinfo {author} {\bibfnamefont {M.~B.}\
  \bibnamefont {Schilling}}, \bibinfo {author} {\bibfnamefont {C.}~\bibnamefont
  {Shekhar}}, \bibinfo {author} {\bibfnamefont {C.}~\bibnamefont {Felser}},
  \bibinfo {author} {\bibfnamefont {M.}~\bibnamefont {Dressel}},\ and\ \bibinfo
  {author} {\bibfnamefont {A.~V.}\ \bibnamefont {Pronin}},\ }\bibfield  {title}
  {\bibinfo {title} {{Optical conductivity of the Weyl semimetal NbP}},\ }\href
  {https://doi.org/10.1103/PhysRevB.98.195203} {\bibfield  {journal} {\bibinfo
  {journal} {Phys. Rev. B}\ }\textbf {\bibinfo {volume} {98}},\ \bibinfo
  {pages} {195203} (\bibinfo {year} {2018})}\BibitemShut {NoStop}%
\bibitem [{\citenamefont {H\"utt}\ \emph {et~al.}(2018)\citenamefont {H\"utt},
  \citenamefont {Yaresko}, \citenamefont {Schilling}, \citenamefont {Shekhar},
  \citenamefont {Felser}, \citenamefont {Dressel},\ and\ \citenamefont
  {Pronin}}]{Huett2018}%
  \BibitemOpen
  \bibfield  {author} {\bibinfo {author} {\bibfnamefont {F.}~\bibnamefont
  {H\"utt}}, \bibinfo {author} {\bibfnamefont {A.}~\bibnamefont {Yaresko}},
  \bibinfo {author} {\bibfnamefont {M.~B.}\ \bibnamefont {Schilling}}, \bibinfo
  {author} {\bibfnamefont {C.}~\bibnamefont {Shekhar}}, \bibinfo {author}
  {\bibfnamefont {C.}~\bibnamefont {Felser}}, \bibinfo {author} {\bibfnamefont
  {M.}~\bibnamefont {Dressel}},\ and\ \bibinfo {author} {\bibfnamefont {A.~V.}\
  \bibnamefont {Pronin}},\ }\bibfield  {title} {\bibinfo {title}
  {{Linear-in-Frequency Optical Conductivity in GdPtBi due to Transitions near
  the Triple Points}},\ }\href {https://doi.org/10.1103/PhysRevLett.121.176601}
  {\bibfield  {journal} {\bibinfo  {journal} {Phys. Rev. Lett.}\ }\textbf
  {\bibinfo {volume} {121}},\ \bibinfo {pages} {176601} (\bibinfo {year}
  {2018})}\BibitemShut {NoStop}%
\bibitem [{\citenamefont {Carbotte}(2016)}]{Carbotte2016}%
  \BibitemOpen
  \bibfield  {author} {\bibinfo {author} {\bibfnamefont {J.~P.}\ \bibnamefont
  {Carbotte}},\ }\bibfield  {title} {\bibinfo {title} {{Dirac cone tilt on
  interband optical background of type-I and type-II Weyl semimetals}},\ }\href
  {https://doi.org/10.1103/PhysRevB.94.165111} {\bibfield  {journal} {\bibinfo
  {journal} {Phys. Rev. B}\ }\textbf {\bibinfo {volume} {94}},\ \bibinfo
  {pages} {165111} (\bibinfo {year} {2016})}\BibitemShut {NoStop}%
\bibitem [{\citenamefont {Jiang}\ \emph {et~al.}(2017)\citenamefont {Jiang},
  \citenamefont {Liu}, \citenamefont {Sun}, \citenamefont {Yang}, \citenamefont
  {Rajamathi}, \citenamefont {Qi}, \citenamefont {Yang}, \citenamefont {Chen},
  \citenamefont {Peng}, \citenamefont {Hwang}, \citenamefont {Sun},
  \citenamefont {Mo}, \citenamefont {Vobornik}, \citenamefont {Fujii},
  \citenamefont {Parkin}, \citenamefont {Felser}, \citenamefont {Yan},\ and\
  \citenamefont {Chen}}]{Jiang2017}%
  \BibitemOpen
  \bibfield  {author} {\bibinfo {author} {\bibfnamefont {J.}~\bibnamefont
  {Jiang}}, \bibinfo {author} {\bibfnamefont {Z.~K.}\ \bibnamefont {Liu}},
  \bibinfo {author} {\bibfnamefont {Y.}~\bibnamefont {Sun}}, \bibinfo {author}
  {\bibfnamefont {H.~F.}\ \bibnamefont {Yang}}, \bibinfo {author}
  {\bibfnamefont {C.~R.}\ \bibnamefont {Rajamathi}}, \bibinfo {author}
  {\bibfnamefont {Y.~P.}\ \bibnamefont {Qi}}, \bibinfo {author} {\bibfnamefont
  {L.~X.}\ \bibnamefont {Yang}}, \bibinfo {author} {\bibfnamefont
  {C.}~\bibnamefont {Chen}}, \bibinfo {author} {\bibfnamefont {H.}~\bibnamefont
  {Peng}}, \bibinfo {author} {\bibfnamefont {C.-C.}\ \bibnamefont {Hwang}},
  \bibinfo {author} {\bibfnamefont {S.~Z.}\ \bibnamefont {Sun}}, \bibinfo
  {author} {\bibfnamefont {S.-K.}\ \bibnamefont {Mo}}, \bibinfo {author}
  {\bibfnamefont {I.}~\bibnamefont {Vobornik}}, \bibinfo {author}
  {\bibfnamefont {J.}~\bibnamefont {Fujii}}, \bibinfo {author} {\bibfnamefont
  {S.~S.~P.}\ \bibnamefont {Parkin}}, \bibinfo {author} {\bibfnamefont
  {C.}~\bibnamefont {Felser}}, \bibinfo {author} {\bibfnamefont {B.~H.}\
  \bibnamefont {Yan}},\ and\ \bibinfo {author} {\bibfnamefont {Y.~L.}\
  \bibnamefont {Chen}},\ }\bibfield  {title} {\bibinfo {title} {{Signature of
  type-II Weyl semimetal phase in MoTe2}},\ }\href
  {https://doi.org/https://doi.org/10.1038/ncomms13973} {\bibfield  {journal}
  {\bibinfo  {journal} {Nature Communications}\ }\textbf {\bibinfo {volume}
  {8}},\ \bibinfo {pages} {13973} (\bibinfo {year} {2017})}\BibitemShut
  {NoStop}%
\bibitem [{\citenamefont {Santos-Cottin}\ \emph {et~al.}(2019)\citenamefont
  {Santos-Cottin}, \citenamefont {Martino}, \citenamefont {Mardele},
  \citenamefont {Witteveen}, \citenamefont {von Rohr}, \citenamefont {Homes},
  \citenamefont {Rukelj},\ and\ \citenamefont {Akrap}}]{santos2019}%
  \BibitemOpen
  \bibfield  {author} {\bibinfo {author} {\bibfnamefont {D.}~\bibnamefont
  {Santos-Cottin}}, \bibinfo {author} {\bibfnamefont {E.}~\bibnamefont
  {Martino}}, \bibinfo {author} {\bibfnamefont {F.~L.}\ \bibnamefont
  {Mardele}}, \bibinfo {author} {\bibfnamefont {C.}~\bibnamefont {Witteveen}},
  \bibinfo {author} {\bibfnamefont {F.~O.}\ \bibnamefont {von Rohr}}, \bibinfo
  {author} {\bibfnamefont {C.~C.}\ \bibnamefont {Homes}}, \bibinfo {author}
  {\bibfnamefont {Z.}~\bibnamefont {Rukelj}},\ and\ \bibinfo {author}
  {\bibfnamefont {A.}~\bibnamefont {Akrap}},\ }\href@noop {} {} (\bibinfo
  {year} {2019}),\ \Eprint {https://arxiv.org/abs/1910.05088}
  {arXiv:1910.05088} \BibitemShut {NoStop}%
\bibitem [{\citenamefont {Kimura}\ \emph {et~al.}(2019)\citenamefont {Kimura},
  \citenamefont {Nakajima}, \citenamefont {Mita}, \citenamefont {Jha},
  \citenamefont {Higashinaka}, \citenamefont {Matsuda},\ and\ \citenamefont
  {Aoki}}]{Kimura2019}%
  \BibitemOpen
  \bibfield  {author} {\bibinfo {author} {\bibfnamefont {S.-i.}\ \bibnamefont
  {Kimura}}, \bibinfo {author} {\bibfnamefont {Y.}~\bibnamefont {Nakajima}},
  \bibinfo {author} {\bibfnamefont {Z.}~\bibnamefont {Mita}}, \bibinfo {author}
  {\bibfnamefont {R.}~\bibnamefont {Jha}}, \bibinfo {author} {\bibfnamefont
  {R.}~\bibnamefont {Higashinaka}}, \bibinfo {author} {\bibfnamefont {T.~D.}\
  \bibnamefont {Matsuda}},\ and\ \bibinfo {author} {\bibfnamefont
  {Y.}~\bibnamefont {Aoki}},\ }\bibfield  {title} {\bibinfo {title} {{Optical
  evidence of the type-II Weyl semimetals ${\mathrm{MoTe}}_{2}$ and
  ${\mathrm{WTe}}_{2}$}},\ }\href {https://doi.org/10.1103/PhysRevB.99.195203}
  {\bibfield  {journal} {\bibinfo  {journal} {Phys. Rev. B}\ }\textbf {\bibinfo
  {volume} {99}},\ \bibinfo {pages} {195203} (\bibinfo {year}
  {2019})}\BibitemShut {NoStop}%
\bibitem [{\citenamefont {Homes}\ \emph {et~al.}(2015)\citenamefont {Homes},
  \citenamefont {Ali},\ and\ \citenamefont {Cava}}]{Homes2015}%
  \BibitemOpen
  \bibfield  {author} {\bibinfo {author} {\bibfnamefont {C.~C.}\ \bibnamefont
  {Homes}}, \bibinfo {author} {\bibfnamefont {M.~N.}\ \bibnamefont {Ali}},\
  and\ \bibinfo {author} {\bibfnamefont {R.~J.}\ \bibnamefont {Cava}},\
  }\bibfield  {title} {\bibinfo {title} {{Optical properties of the perfectly
  compensated semimetal ${\mathrm{WTe}}_{2}$}},\ }\href
  {https://doi.org/10.1103/PhysRevB.92.161109} {\bibfield  {journal} {\bibinfo
  {journal} {Phys. Rev. B}\ }\textbf {\bibinfo {volume} {92}},\ \bibinfo
  {pages} {161109} (\bibinfo {year} {2015})}\BibitemShut {NoStop}%
\bibitem [{\citenamefont {Wang}\ \emph {et~al.}(2016)\citenamefont {Wang},
  \citenamefont {Zhang}, \citenamefont {Huang}, \citenamefont {Nie},
  \citenamefont {Liu}, \citenamefont {Liang}, \citenamefont {Zhang},
  \citenamefont {Shen}, \citenamefont {Liu}, \citenamefont {Hu}, \citenamefont
  {Ding}, \citenamefont {Liu}, \citenamefont {Hu}, \citenamefont {He},
  \citenamefont {Zhao}, \citenamefont {Yu}, \citenamefont {Hu}, \citenamefont
  {Wei}, \citenamefont {Mao}, \citenamefont {Shi}, \citenamefont {Jia},
  \citenamefont {Zhang}, \citenamefont {Zhang}, \citenamefont {Yang},
  \citenamefont {Wang}, \citenamefont {Peng}, \citenamefont {Weng},
  \citenamefont {Dai}, \citenamefont {Fang}, \citenamefont {Xu}, \citenamefont
  {Chen},\ and\ \citenamefont {Zhou}}]{Wang2016}%
  \BibitemOpen
  \bibfield  {author} {\bibinfo {author} {\bibfnamefont {C.}~\bibnamefont
  {Wang}}, \bibinfo {author} {\bibfnamefont {Y.}~\bibnamefont {Zhang}},
  \bibinfo {author} {\bibfnamefont {J.}~\bibnamefont {Huang}}, \bibinfo
  {author} {\bibfnamefont {S.}~\bibnamefont {Nie}}, \bibinfo {author}
  {\bibfnamefont {G.}~\bibnamefont {Liu}}, \bibinfo {author} {\bibfnamefont
  {A.}~\bibnamefont {Liang}}, \bibinfo {author} {\bibfnamefont
  {Y.}~\bibnamefont {Zhang}}, \bibinfo {author} {\bibfnamefont
  {B.}~\bibnamefont {Shen}}, \bibinfo {author} {\bibfnamefont {J.}~\bibnamefont
  {Liu}}, \bibinfo {author} {\bibfnamefont {C.}~\bibnamefont {Hu}}, \bibinfo
  {author} {\bibfnamefont {Y.}~\bibnamefont {Ding}}, \bibinfo {author}
  {\bibfnamefont {D.}~\bibnamefont {Liu}}, \bibinfo {author} {\bibfnamefont
  {Y.}~\bibnamefont {Hu}}, \bibinfo {author} {\bibfnamefont {S.}~\bibnamefont
  {He}}, \bibinfo {author} {\bibfnamefont {L.}~\bibnamefont {Zhao}}, \bibinfo
  {author} {\bibfnamefont {L.}~\bibnamefont {Yu}}, \bibinfo {author}
  {\bibfnamefont {J.}~\bibnamefont {Hu}}, \bibinfo {author} {\bibfnamefont
  {J.}~\bibnamefont {Wei}}, \bibinfo {author} {\bibfnamefont {Z.}~\bibnamefont
  {Mao}}, \bibinfo {author} {\bibfnamefont {Y.}~\bibnamefont {Shi}}, \bibinfo
  {author} {\bibfnamefont {X.}~\bibnamefont {Jia}}, \bibinfo {author}
  {\bibfnamefont {F.}~\bibnamefont {Zhang}}, \bibinfo {author} {\bibfnamefont
  {S.}~\bibnamefont {Zhang}}, \bibinfo {author} {\bibfnamefont
  {F.}~\bibnamefont {Yang}}, \bibinfo {author} {\bibfnamefont {Z.}~\bibnamefont
  {Wang}}, \bibinfo {author} {\bibfnamefont {Q.}~\bibnamefont {Peng}}, \bibinfo
  {author} {\bibfnamefont {H.}~\bibnamefont {Weng}}, \bibinfo {author}
  {\bibfnamefont {X.}~\bibnamefont {Dai}}, \bibinfo {author} {\bibfnamefont
  {Z.}~\bibnamefont {Fang}}, \bibinfo {author} {\bibfnamefont {Z.}~\bibnamefont
  {Xu}}, \bibinfo {author} {\bibfnamefont {C.}~\bibnamefont {Chen}},\ and\
  \bibinfo {author} {\bibfnamefont {X.~J.}\ \bibnamefont {Zhou}},\ }\bibfield
  {title} {\bibinfo {title} {{Observation of Fermi arc and its connection with
  bulk states in the candidate type-II Weyl semimetal ${\mathrm{WTe}}_{2}$}},\
  }\href {https://doi.org/https://doi.org/10.1103/PhysRevB.94.121113}
  {\bibfield  {journal} {\bibinfo  {journal} {Phys. Rev. B}\ }\textbf {\bibinfo
  {volume} {94}},\ \bibinfo {pages} {241119} (\bibinfo {year}
  {2016})}\BibitemShut {NoStop}%
\bibitem [{\citenamefont {Frenzel}\ \emph {et~al.}(2017)\citenamefont
  {Frenzel}, \citenamefont {Homes}, \citenamefont {Gibson}, \citenamefont
  {Shao}, \citenamefont {Post}, \citenamefont {Charnukha}, \citenamefont
  {Cava},\ and\ \citenamefont {Basov}}]{Frenzel2017}%
  \BibitemOpen
  \bibfield  {author} {\bibinfo {author} {\bibfnamefont {A.~J.}\ \bibnamefont
  {Frenzel}}, \bibinfo {author} {\bibfnamefont {C.~C.}\ \bibnamefont {Homes}},
  \bibinfo {author} {\bibfnamefont {Q.~D.}\ \bibnamefont {Gibson}}, \bibinfo
  {author} {\bibfnamefont {Y.~M.}\ \bibnamefont {Shao}}, \bibinfo {author}
  {\bibfnamefont {K.~W.}\ \bibnamefont {Post}}, \bibinfo {author}
  {\bibfnamefont {A.}~\bibnamefont {Charnukha}}, \bibinfo {author}
  {\bibfnamefont {R.~J.}\ \bibnamefont {Cava}},\ and\ \bibinfo {author}
  {\bibfnamefont {D.~N.}\ \bibnamefont {Basov}},\ }\bibfield  {title} {\bibinfo
  {title} {{Anisotropic electrodynamics of type-II Weyl semimetal candidate
  ${\mathrm{WTe}}_{2}$}},\ }\href {https://doi.org/10.1103/PhysRevB.95.245140}
  {\bibfield  {journal} {\bibinfo  {journal} {Phys. Rev. B}\ }\textbf {\bibinfo
  {volume} {95}},\ \bibinfo {pages} {245140} (\bibinfo {year}
  {2017})}\BibitemShut {NoStop}%
\bibitem [{\citenamefont {Xi}\ \emph {et~al.}(2013)\citenamefont {Xi},
  \citenamefont {Ma}, \citenamefont {Liu}, \citenamefont {Chen}, \citenamefont
  {Ku}, \citenamefont {Berger}, \citenamefont {Martin}, \citenamefont
  {Tanner},\ and\ \citenamefont {Carr}}]{Xi2013}%
  \BibitemOpen
  \bibfield  {author} {\bibinfo {author} {\bibfnamefont {X.}~\bibnamefont
  {Xi}}, \bibinfo {author} {\bibfnamefont {C.}~\bibnamefont {Ma}}, \bibinfo
  {author} {\bibfnamefont {Z.}~\bibnamefont {Liu}}, \bibinfo {author}
  {\bibfnamefont {Z.}~\bibnamefont {Chen}}, \bibinfo {author} {\bibfnamefont
  {W.}~\bibnamefont {Ku}}, \bibinfo {author} {\bibfnamefont {H.}~\bibnamefont
  {Berger}}, \bibinfo {author} {\bibfnamefont {C.}~\bibnamefont {Martin}},
  \bibinfo {author} {\bibfnamefont {D.~B.}\ \bibnamefont {Tanner}},\ and\
  \bibinfo {author} {\bibfnamefont {G.~L.}\ \bibnamefont {Carr}},\ }\bibfield
  {title} {\bibinfo {title} {{Signatures of a Pressure-Induced Topological
  Quantum Phase Transition in BiTeI}},\ }\href
  {https://doi.org/https://doi.org/10.1103/PhysRevLett.111.155701} {\bibfield
  {journal} {\bibinfo  {journal} {Phys. Rev. Lett.}\ }\textbf {\bibinfo
  {volume} {111}},\ \bibinfo {pages} {155701} (\bibinfo {year}
  {2013})}\BibitemShut {NoStop}%
\bibitem [{\citenamefont {Tran}\ \emph {et~al.}(2014)\citenamefont {Tran},
  \citenamefont {Levallois}, \citenamefont {Lerch}, \citenamefont {Teyssier},
  \citenamefont {Kuzmenko}, \citenamefont {Aut\`es}, \citenamefont {Yazyev},
  \citenamefont {Ubaldini}, \citenamefont {Giannini}, \citenamefont {van~der
  Marel},\ and\ \citenamefont {Akrap}}]{Tran2014}%
  \BibitemOpen
  \bibfield  {author} {\bibinfo {author} {\bibfnamefont {M.~K.}\ \bibnamefont
  {Tran}}, \bibinfo {author} {\bibfnamefont {J.}~\bibnamefont {Levallois}},
  \bibinfo {author} {\bibfnamefont {P.}~\bibnamefont {Lerch}}, \bibinfo
  {author} {\bibfnamefont {J.}~\bibnamefont {Teyssier}}, \bibinfo {author}
  {\bibfnamefont {A.~B.}\ \bibnamefont {Kuzmenko}}, \bibinfo {author}
  {\bibfnamefont {G.}~\bibnamefont {Aut\`es}}, \bibinfo {author} {\bibfnamefont
  {O.~V.}\ \bibnamefont {Yazyev}}, \bibinfo {author} {\bibfnamefont
  {A.}~\bibnamefont {Ubaldini}}, \bibinfo {author} {\bibfnamefont
  {E.}~\bibnamefont {Giannini}}, \bibinfo {author} {\bibfnamefont
  {D.}~\bibnamefont {van~der Marel}},\ and\ \bibinfo {author} {\bibfnamefont
  {A.}~\bibnamefont {Akrap}},\ }\bibfield  {title} {\bibinfo {title}
  {{Infrared- and Raman-Spectroscopy Measurements of a Transition in the
  Crystal Structure and a Closing of the Energy Gap of BiTeI under Pressure}},\
  }\href {https://doi.org/https://doi.org/10.1103/PhysRevLett.112.047402}
  {\bibfield  {journal} {\bibinfo  {journal} {Phys. Rev. Lett.}\ }\textbf
  {\bibinfo {volume} {112}},\ \bibinfo {pages} {047402} (\bibinfo {year}
  {2014})}\BibitemShut {NoStop}%
\end{thebibliography}%

\makeatletter
\renewcommand\@bibitem[1]{\item\if@filesw \immediate\write\@auxout
    {\string\bibcite{#1}{S\the\value{\@listctr}}}\fi\ignorespaces}% <------------
\def\@biblabel#1{[S#1]}% <-------------------
\makeatother
\setcounter{figure}{0}
\renewcommand{\thefigure}{S\arabic{figure}}
\newcommand{\av}{\mathbf a}
\newcommand{\bv}{\mathbf b}
\newcommand{\cv}{\mathbf c}

\newpage
\begin{center}
\textbf{Supplemental Material for\\ ``Two linear regimes in optical conductivity of a Type-I Weyl semimetal: the case of elemental tellurium"}
\end{center}
D. Rodriguez, A. A. Tsirlin, T. Biesner, T. Ueno, T. Takahashi, K. Kobayashi, M. Dressel, E. Uykur

\subsection{Samples and infrared measurements}

Single crystals of Tellurium are prepared by the Bridgman method. The crystals are easily cleaved at liquid nitrogen temperature perpendicular to the $b$-axis. Room-temperature reflectivity measurements have been performed on $ac$-plane of a single crystal with the typical size of 160~$\mu$m$\times$170~$\mu$m$\times$60~$\mu$m. Owing to the small sample size we did not utilized polarizers and the observed properties reflect the average over the $ac$-plane.  

\begin{figure} [h]
\centering
\includegraphics[width=1\linewidth]{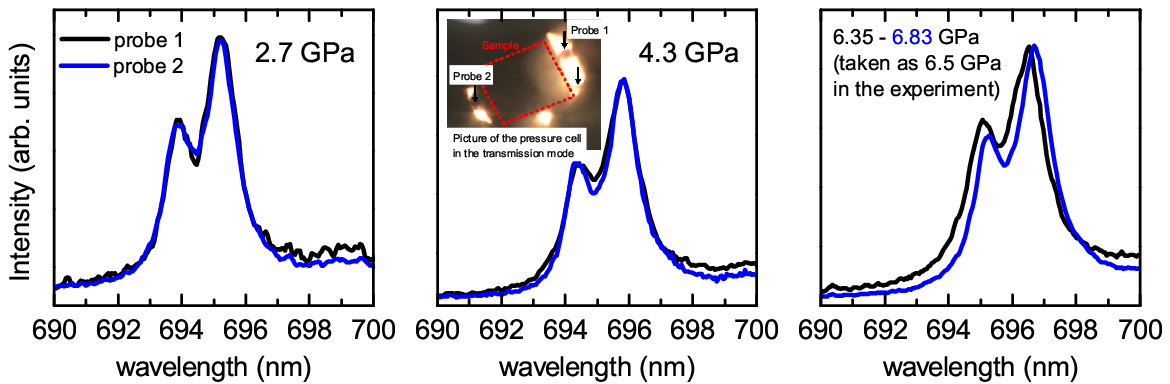}%
\caption{ Ruby luminescence spectra for two different ruby spheres placed inside the pressure cell for various pressures. }
\label{ruby}%
\end{figure}

High pressures in reflectivity measurements were generated with a screw driven diamond anvil cell (DAC). The type-IIa diamond anvils with culet diameter of 800~$\mu$m allow us to reach the pressure range up to 8.44~GPa. A 250 $\mu$m diameter hole is drilled on a CuBe gasket preindented to a thickness of 80 $\mu$m. This hole is used as a sample chamber, where several ruby spheres were loaded together with the sample. The determination of the pressure inside the cell was achieved by monitoring the calibrated shift of the ruby R1 fluorescence line \hyperlink{thesentences}{\color{blue}[S1]}. The DAC was filled with CsI powder as a quasi-hydrostatic pressure-transmitting medium keeping a clean diamond-sample interface. The pressure gradient in the cell is monitored during the measurements via two different Ruby spheres sitting at the opposite sides of the sample. At low pressures (below 5-6~GPa), no pressure gradient has been observed within the accuracy of the CCD spectrometer used in Ruby luminescence measurements (0.05~nm / 0.15~GPa). At high pressures a gradient up to $\sim$0.7 GPa is present [Fig.~\ref{ruby}]. 

\begin{figure}
\centering
\includegraphics[width=1\linewidth]{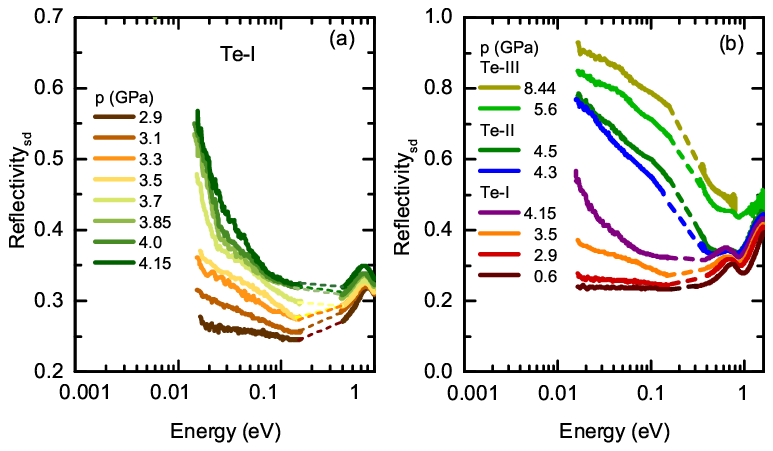}%
\caption{ Measured reflectivity (a) in the low-pressure Te-I phase and (b) through the structural phase transitions. Dashed lines are the energy range, where the two-phonon absorption of the diamond anvil cell is dominant. In this energy range, a linear extrapolation have been used for the further analysis.  }
\label{OptCon}%
\end{figure}

Measurements have been performed with a Hyperion infrared microscope coupled to a Bruker Vertex 80v Fourier transform infrared spectrometer. The pressure cell is attached to the infrared microscope employing a custom made setup to control the xyz-position and the rotation of the pressure cell. Reflectivity spectra at the sample-diamond interface, R$_{sd}$ were collected between 100 and 20000 \cm and the CuBe gasket was used as a reference. The measured reflectivity spectra for the selected pressures have been given in Fig.~\ref{OptCon}. Here, the dashed lines show the energy range, where the spectra are affected by the multi-phonon absorption of the diamond anvil cell, where a linear extrapolation have been used for the further analysis. 
The real part of the optical conductivity, $\sigma_1$,  was extracted via Kramers-Kronig (KK) transformations considering sample-diamond interface parameters during the calculations \hyperlink{thesentences}{\color{blue}[S2]}. Simultaneous fits of reflectivity and the optical conductivity by the Drude-Lorentz model ensure that the KK procedure has been carried out correctly.

\subsection{Determination of the optical conductivity}

In our high-pressure infrared spectroscopy technique, the reflectivity spectrum of tellurium $R_{sd}(\omega)$ is measured at the sample-diamond interface in the diamond anvil cell (DAC) which is expressed as 
\begin{equation}
R_{sd}(\omega)=R_{CuBe-diamond} \frac{I_{sd}(\omega)}{I_{CuBe-diamond}(\omega)}
\label{reflectivity}
\end{equation}
Here $I_{sd}(\omega)$  is the reflected light intensity at the sample-diamond interface and $I_{CuBe-diamond}(\omega)$ is at the gasket-diamond interface. At each pressure the reflectivity of the sample has been corrected for the gasket reflection with the $R_{CuBe-diamond}$. Thus, Eq.~\ref{reflectivity} gives us the absolute reflectivity $R_{sd}(\omega)$ obtained at the sample-diamond interface. \\

During the KK analysis of the reflectivity measured at the sample-diamond interface, the standard relation between the phase and the reflectivity needs to be corrected, which gives an additional term (second term in the right hand side of Eq.~\ref{KK}). Namely, the presence of a medium with the refractive index $>$1 (in the case of diamond it is 2.38) brings an extra phase shift.  

\begin{equation}
\theta(\omega_0) =- \frac{\omega_0}{\pi}P \int_0^{+\infty} \frac{\ln R_{sd}(\omega)}{\omega^2-\omega_0^2}\, d\omega+\left[\pi-2\text{arctan}\frac{\omega_{\beta}}{\omega_0}\right]
\label{KK}
\end{equation}

$P$ is the principal value of the complex response function. $\omega_{\beta}$ is the position of the reflectivity pole on the imaginary axis of the complex plane. For the case of measurements performed at the sample-vacuum interface, $\omega_{\beta}$ tends to go to infinity eliminating the second term and leading to the common KK analysis.\\

$\omega_{\beta}$ is $a$ $priori$ an unknown parameter, however, it can be estimated by comparing the lowest-pressure optical conductivity (in our case $<$0.2~GPa spectrum is used) with the ambient-pressure one, which can be obtained through the regular KK analysis. We found the best fitting for the parameter $\omega_{\beta}$ = 25.000 cm$^{-1}$. 

\subsection{Determination of fit parameters}

\begin{figure} [h]
\centering
\includegraphics[width=1\linewidth]{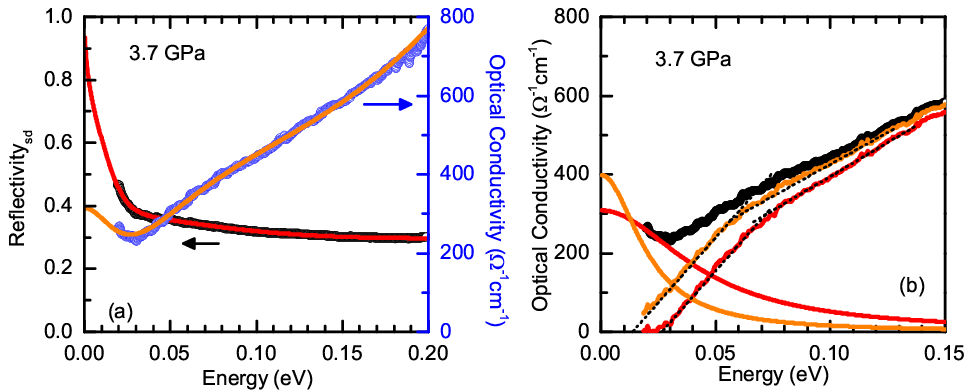}%
\caption{ (a) Simultaneous fits of reflectivity and optical conductivity at 3.7~GPa. (b) The effect of choosing different Drude contributions on to the interband transitions. The kink behavior and the extrapolation of the low energy optical conductivity to the finite energies are robust under different choices. }
\label{robust}%
\end{figure}

Since the DFT calculations do not include intraband transitions, these contributions should be subtracted from the experimental curves for a direct comparison. To identify the Drude-like contribution, we performed Drude-Lorentz fits to the reflectivity and optical conductivity, simultaneously. These fits restrain the parameters, because those should satisfy the real and the imaginary part of the dielectric function, simultaneously. An example of this fit is shown in Fig.~\ref{robust}(a) for 3.7~GPa.     

On the other hand, with several absorption bands present, the final fits are not necessarily the unique solutions and some spectral weight redistribution is possible between different contributions. In Fig.~\ref{robust}(b), we showed two of the extreme cases for the intraband transitions, which in principle can give a similar reflectivity and optical conductivity fit, but leads to a slightly different interband transitions. Please note that the discussed kink-behavior and the extrapolation of the low energy optical conductivity to the finite frequencies with increasing pressure are robust features. In the calculated optical conductivity spectra, these features correlate with the slope change of the optical conductivity (kink behavior) and the small upturn of the optical conductivity (extrapolation to the finite frequencies). The choice of the intraband contribution only affects the energy range slightly, while the features are still visible. 

To keep the consistency throughout the measured pressure range, we choose the Drude-component with the largest $1/\tau$, because this can be determined in a same manner with increasing pressure allowing us to discuss the qualitative pressure evolution of the features.  

\subsection{Computational methods}

Full-relativistic density-functional (DFT) calculations were performed within the Wien2K code~\hyperlink{thesentences}{\color{blue}[S3]} using the \texttt{optic} module~\hyperlink{thesentences}{\color{blue}[S4]} for evaluating optical conductivity. The plane-wave cutoff of RKMAX=7 was used. We employed the modified Becke-Johnson (mBJ) exchange-correlation functional~\hyperlink{thesentences}{\color{blue}[S5]}, because it delivers a realistic band gap of Te-I at ambient pressure, whereas conventional local-density (LDA) or generalized-gradient (GGA) approximations render Te-I metallic already at ambient pressure when relativistic effects are included. This strategy is similar to earlier computational studies of Te-I~\hyperlink{thesentences}{\color{blue}[S6]}. The parametrization from Ref.~\hyperlink{thesentences}{\color{blue}[S7]} was used in mBJ~\hyperlink{thesentences}{\color{blue}[S8]}. We note in passing that the mBJ functional has not been parametrized for semiconductors with very small band gaps and may not perfectly reproduce the critical pressure of the semiconductor-to-metal transition in Te-I. Indeed, our calculations suggest metallic band structure at 1.88\,GPa, whereas experimentally Te-I may not be metallic until about 3\,GPa. This discrepancy is, however, unavoidable with virtually any \textit{ab initio} method, and arguably unimportant, because calculations still reproduce band structures of both semiconducting and metallic Te-I, albeit with a slightly shifted pressure scale.

\begin{figure}
\centering
\includegraphics[width=1\linewidth]{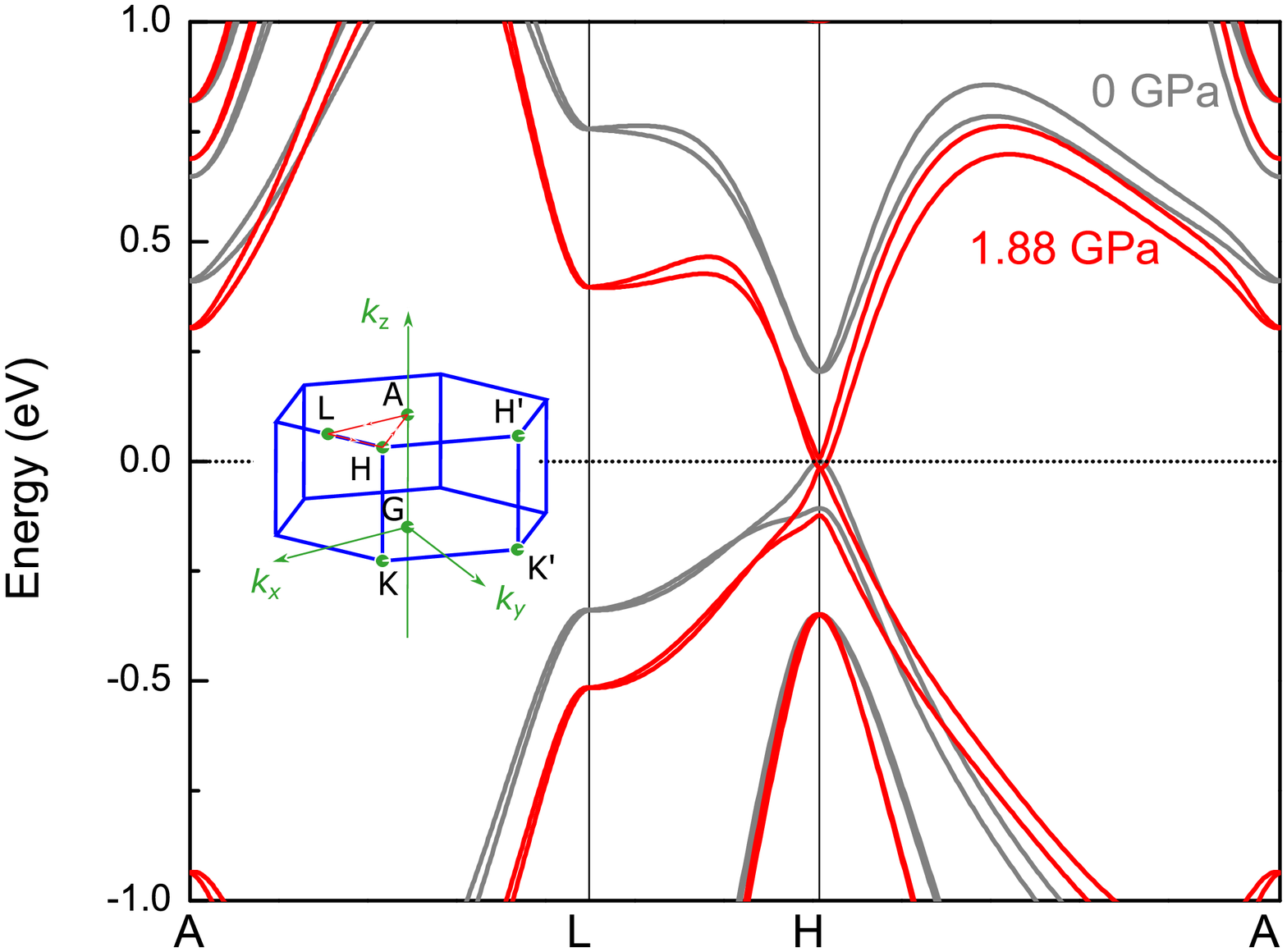}%
\caption{Band structure of Te-I at ambient pressure and 1.88~GPa, where the bands are just touching, along the $k$-path indicated with the red line in Brillouin zone sketch.}
\label{TeIbands}%
\end{figure}

All calculations were performed for the experimental crystal structures, because each DFT functional has its own bias toward metallic or nonmetallic states, and \textit{ab initio} structure relaxation will bias the results accordingly. In the case of Te-I, we used the experimental structural parameters from Ref.~\hyperlink{thesentences}{\color{blue}[S9]}, which also determines the pressure values used in our calculations. Band dispersions [Fig.~\ref{TeIbands}] were calculated on the well-converged $48\times 48\times 48$ $k$-mesh and for optical conductivity we used denser meshes with up to $140\times 140\times 140$ points. Optical conductivity given as $\sigma_1(\omega)$ is determined by taking the average of $\sigma_{xx}$ and $\sigma_{zz}$ ($\sigma_1(\omega) = (\sigma_{xx} + \sigma_{zz})/2$), reflecting the measurement configuration. 

\subsection{Comparison of the other bands}

\begin{figure*}
\centering
\includegraphics[width=1\linewidth]{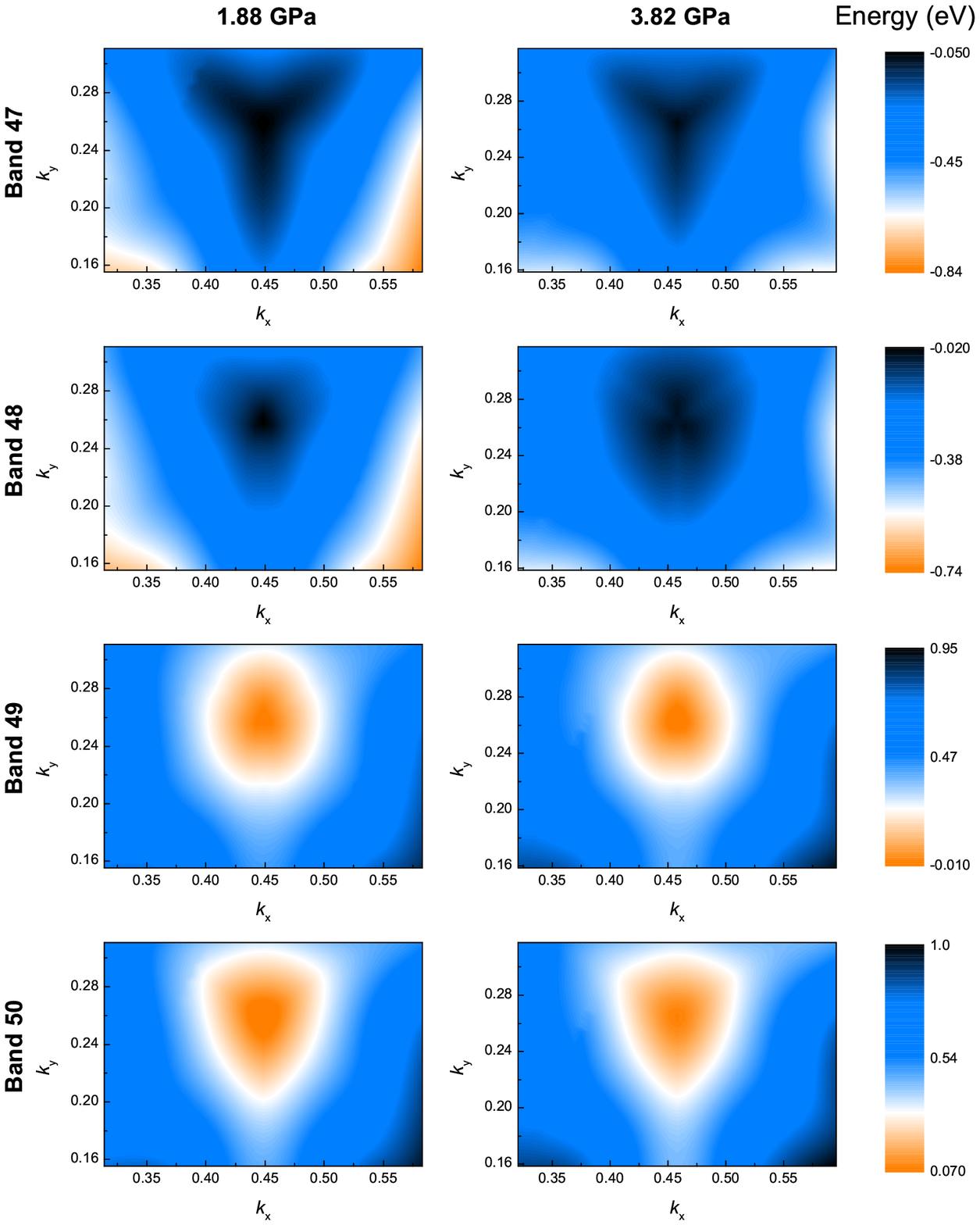}%
\caption{2D profiles of bands (Band 47, 48, 49, and 50) around $H$ point giving rise to significant contribution to the low energy optical conductivity. Left pannels: 1.88~GPa and Right pannels: 3.82~GPa.}
\label{OtherBands}%
\end{figure*}

In the main text, the comparison of the 2D profile for the band 48 between 1.88 and 3.82 GPa has been given. This band is shown to change its profile with external pressure. Here in the supplementary we plotted the 2D profile of the other bands around $H$ point, which contribute to the optical conductivity significantly [Fig.~\ref{OtherBands}]. Left pannels are correspond to the 1.88~GPa, while right pannels show the bands at 3.82~GPa. Each row belong to one significant band as stated at the left handside of the plots. 

\newpage
\makeatletter
\renewcommand\@bibitem[1]{\item\if@filesw \immediate\write\@auxout
    {\string\bibcite{#1}{S\the\value{\@listctr}}}\fi\ignorespaces}% <------------
\def\@biblabel#1{[S#1]}% <-------------------
\makeatother

\section*{Supplementary references}
\begin{small}

\setlength\parindent{0pt}\hypertarget{thesentences}[{S1}] H. K. Mao, J. Xu, and P. M. Bell, ``Calibration of the ruby pressure gauge to 800 kbar under quasi-hydrostatic conditions," \href{https://doi.org/10.1029/JB091iB05p04673} {J. Geophys. Res. \textbf{91,} 4673 (1986)}.\\

\setlength\parindent{0pt}[S2] A. Pashkin, M. Dressel, and C. A. Kuntscher, ``Pressure-induced deconfinement of the charge transport in the quasi-one-dimensional mott insulator TMTTF$_2$AsF$_6$," \href{https://doi.org/10.1103/PhysRevB.74.165118}{Phys. Rev. B \textbf{74,} 165118 (2006)}.\\

\setlength\parindent{0pt}[S3] P. Blaha, K. Schwarz, G.K.H. Madsen, D. Kvasnicka, J. Luitz, R. Laskowski, F. Tran, and L.D. Marks, WIEN2k, ``An Augmented Plane Wave + Local Orbitals Program for Calculating Crystal Properties," (Karlheinz Schwarz, Techn. Universit\"at Wien, Austria), 2018. ISBN 3-9501031-1-2.\\

\setlength\parindent{0pt}[S4] C. Ambrosch-Draxl and J.O. Sofo. ``Linear optical properties of solids within the full-potential linearized augmented planewave method," \href{https://doi.org/10.1016/j.cpc.2006.03.005} {Comput. Phys. Commun.  \textbf{175}, 1 (2006)}.\\

\setlength\parindent{0pt}[S5] F. Tran and P. Blaha, ``Accurate Band Gaps of Semiconductors and Insulators with a Semilocal Exchange-Correlation Potential," \href{https://doi.org/10.1103/PhysRevLett.102.226401}{Phys. Rev. Lett. \textbf{102}, 226401 (2009)}.\\

\setlength\parindent{0pt}[S6] L.A. Agapito, N. Kioussis, W.A. Goddard, and N.P. Ong, ``Novel Family of Chiral-Based Topological Insulators: Elemental Tellurium under Strain,"  \href{https://doi.org/10.1103/PhysRevLett.102.226401}{Phys. Rev. Lett. \textbf{110}, 176401 (2013)}.\\

\setlength\parindent{0pt}[S7] D. Koller, F. Tran, and P. Blaha, ``Improving the modified {Becke-Johnson} exchange potential,"  \href{https://doi.org/10.1103/PhysRevB.85.155109}{Phys. Rev. B \textbf{85}, 155109 (2012)}.\\

\setlength\parindent{0pt}[S8] Specifically, we set $A=0.488$, $B=0.5$, and $c=1$ and also checked that other mBJ parametrizations do not change the results significantly.\\

\setlength\parindent{0pt}[S9] R. Keller, W.B. Holzapfel, and H. Schulz, ``Effect of pressure on the atom positions in {Se} and {Te},"  \href{https://doi.org/10.1103/PhysRevB.16.4404}{Phys. Rev. B \textbf{16}, 4404 (1977)}.\\

\end{small}

\end{document}